# Modulation of Ionic Current Rectification in Short Unipolar Nanopores

Hongwen Zhang,[1, 2] Long Ma,[1, 2] Di Liu,[3] Tianyi Sui,[4] Zuzanna S. Siwy,[5] and Yinghua Qiu[1, 2]*

1. Key Laboratory of High Efficiency and Clean Mechanical Manufacture of the Ministry of Education, State Key Laboratory of Advanced Equipment and Technology for Metal Forming, School of Mechanical Engineering, Shandong University, Jinan, 250061, China

2. Shenzhen Research Institute of Shandong University, Shenzhen, 518000, China

3. Department of Prothodontics, School and Hospital of Stomatology, Cheeloo College of Medicine, Shandong University, Jinan, 250012, China.

4. School of Mechanical Engineering, Tianjin University, Tianjin, 300072, China

5. Department of Physics and Astronomy, University of California, Irvine, California 92697, United States

*Corresponding author: yinghua.qiu@sdu.edu.cn

**Abstract**

With controlled ionic current rectification (ICR) achieved through a strategically designed non-uniform surface charge distribution, short unipolar nanopores exhibit promising applications in nanofluidic sensors, ionic circuits, and ion amplifiers. By systematically investigating how the charged length on inner pore walls modulates ion transport, we found that both the maximum ICR degree and the corresponding charged-length proportion were influenced by nanopore parameters and simulation conditions. For 100 nm-long unipolar nanopores, the highest ICR degree is obtained at a charged-length proportion of ~0.3, due to the corresponding most significant ion enrichment and depletion inside the nanopore under opposite biases. This charged-length proportion of ~0.3 consistently appears as a characteristic value across most considered cases. For short unipolar nanopores, the presence of exterior surface charges significantly enhances the ICR degree by facilitating ion transport through nanopores. The effective widths of charged regions beyond nanopore borders on outer surfaces exhibit direct proportionality to the pore diameter, surface charge density, and applied voltage, and inverse proportionality to the pore length and salt concentration. Our research may provide useful guidance for the design of unipolar nanopores and porous membranes incorporating such charge configurations.

**TOC**

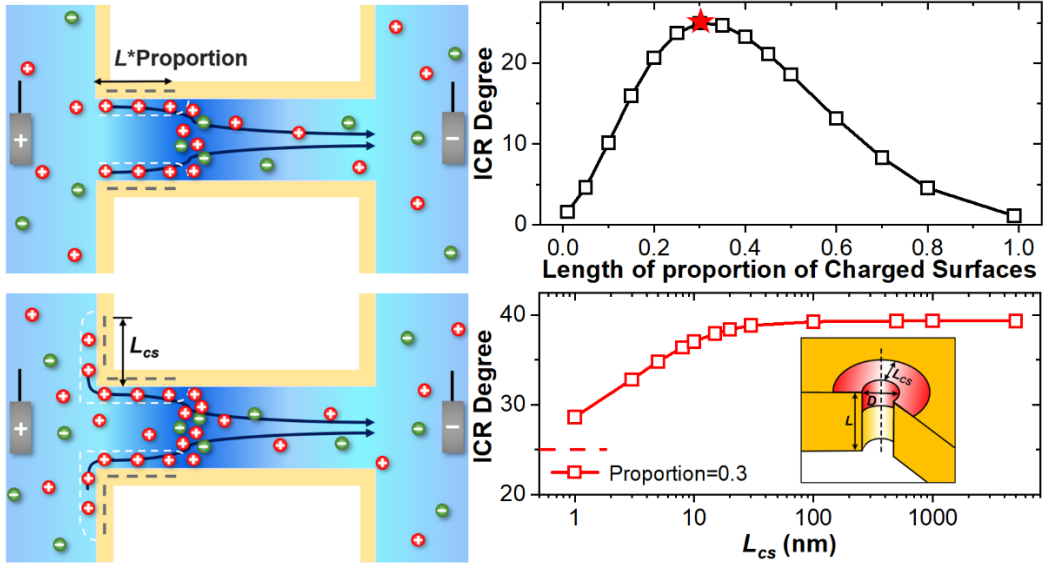

**Introduction**

Under highly confined spaces, the ionic current through nanopores can be influenced significantly by surface charges due to the strong electrostatic interactions with free ions.[1] A series of nanofluidic phenomena have been discovered,[2] including ion concentration polarization,[3] ionic current rectification,[4] and surface-charge-governed ionic conductance,[5] which enable applications in biosensing,[6] iontronic sensors,[7] osmotic energy conversion,[8] desalination,[9] ion separation,[10] and other related fields.[11]

The phenomenon of ionic current rectification (ICR), i.e., different ionic currents through nanopores under equal voltages but opposite biases, reflects a directional preference in ion transport.[12] The emergence of ICR typically results from the asymmetric nanopore shape,[13-15] inhomogeneous charge distributions on pore walls,[16-18] or conductivity gradients across the nanopore.[19-21] The underlying mechanism is generally attributed to the voltage-dependent concentration inside the nanopore, which directly determines the solution conductivity.[22] Analogous to semiconductor diodes, ionic diodes have attracted increasing attention for applications in ion circuits and ion amplifiers.[23-25]

Unipolar nanopores, as a representative type of ionic diodes, rectify ionic current via non-uniform surface charges on pore walls, with distinct charged and uncharged regions. Karnik et al.[16] fabricated the first unipolar diode by introducing a surface charge discontinuity to a channel by diffusion-limited patterning. In their 120-μm-long, 4-μm-wide nanochannels with a height of ~30 nm, significant ICR appeared due to the enrichment and depletion of ions under opposite voltage biases.[17, 26] Later, the ICR phenomenon has been explored with various unipolar nanopores. With track-etched polymer conical pores, Nguyen et al.[27] compared ICR behaviors through both unipolar and bipolar diodes. They found that the bipolar structures produced a higher ICR degree. By modifying nanopores with pH-responsive polymer chains, Ali et al.[28] prepared unipolar diodes with single cigar-shaped nanopores of approximately symmetric geometry, which presented externally tunable nanopore responses. With stepped mesochannels, Yang et al.[29] developed unipolar ionic diodes based on mesoporous silica membranes, demonstrating their potential for osmotic energy conversion. For the detailed illustration of ion transport through unipolar nanopores, various simulations[26, 30] have been conducted to investigate the ICR through unipolar nanopores, which confirm the corresponding ion enrichment and depletion inside the nanopore at the "on" and "off" states.

As a versatile platform, short nanopores offer multiple advantages, including high ion throughput,[31] enhanced detection resolution,[6] and rapid response.[23] As an important electrokinetic phenomenon of ion transport, ICR in ultra-short nanopores has attracted considerable attention.[14, 32-36] However, ICR in ultra-short unipolar nanopores remains rarely explored. Although the ICR mechanism of unipolar nanopores is similar to that of conical nanopores,[37, 38] short conical nanopores always present much lower ICR degrees than those obtained with short unipolar/bipolar nanopores.[14, 30, 33, 39] Also, in practical applications, compared to the fabrication of thin conical nanopores with the asymmetric shape,[40, 41] unipolar charge distribution can be conveniently achieved through surface chemical modification.[42-44] Due to the outstanding performance in unipolar nanopores, the ICR phenomenon in ultra-short unipolar nanopores deserves systematic investigation. Previous studies on bipolar nanopores suggest that the relative length ratio of oppositely charged pore walls significantly affects the ICR performance.[45-47] Considering the vast applications in high-precision sensing, as well as high-performance ion circuits and amplifiers, it is of great importance to build nanofluidic devices with controllable ICR degrees, which requires accurate design of nanopore surface charge properties.

In this work, with COMSOL Multiphysics, the regulation of ion transport characteristics and ICR ratio in short unipolar nanopores was systematically investigated by modulating the charged-length proportion on inner-pore walls. For 100-nm-long unipolar nanopores, the maximum ICR degree occurred at a charged-length proportion of ~0.3, different from the typical value of 0.5 observed in bipolar nanopores. We also explored the variation of the optimal charged-length ratio on the inner-pore walls at the maximum ICR degrees under various nanopore parameters and simulation conditions. The charged-length ratio of ~0.3 persisted as a characteristic proportion in most conditions examined. These findings may provide valuable guidance for the design of unipolar nanopores with controlled ICR performance. Considering that exterior charged surfaces of short nanopores can enhance the ICR degree by providing fast passageways in EDL regions for the transport of counterions, we systematically examined the influence of the exterior charged region on the ICR behavior. The effective charged area on outer surfaces presents direct proportionality to the pore diameter, surface charge density, and applied voltage, inverse proportionality to the pore length and salt concentration, and no dependence on the salt type. These results offer design guidance for unipolar nanopores and porous membranes with tunable ICR properties, for many practical applications such as nanofluidic sensors,[18, 48, 49] ionic logic devices,[50, 51] osmotic energy conversion,[29] and desalination.[52]

**Simulation Methods**

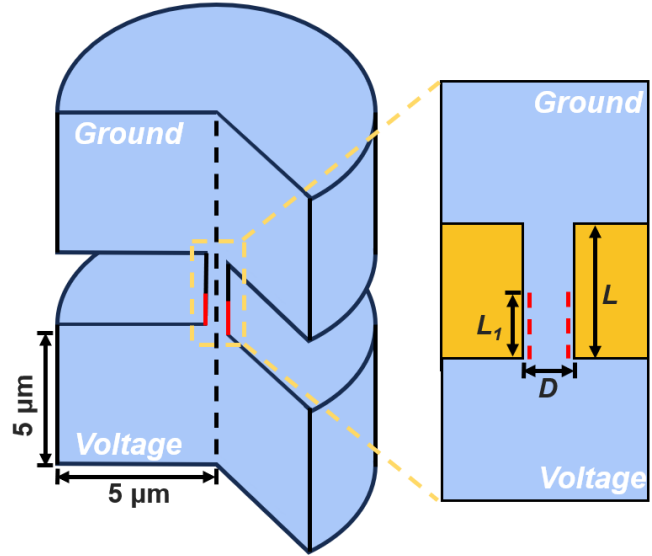

Figure 1 Simulation scheme of ion transport through unipolar nanopores. Two reservoirs with 5 μm in length and 5 μm in radius are located on both sides of the nanopore. External voltages are applied on the boundaries of reservoirs. The pore length and diameter are denoted as $L$ and $D$. On inner pore walls, the length of the charged portion is $L_1$. The charged-length proportion is calculated with $L_1/L$.

Characteristics of ion transport through unipolar nanopores were investigated through COMSOL Multiphysics. As shown in Figure 1, the nanopore connects two reservoirs on both sides.[14, 46, 53] The diameter and length of nanopores are denoted as $D$ and $L$. For unipolar nanopores, a part of the inner surface is charged (length $L_1$) and the rest is neutral. In the simulations, coupled Poisson-Nernst-Planck and Navier-Stokes equations (Eqs. 1-4) were used to describe the distributions of electric potential and ionic concentrations, as well as ion transport and fluid flow under electric fields.[14, 38, 54-56] Detailed boundary conditions are shown in Table S1.

$$\varepsilon\nabla^2\varphi = -\sum_{i=1}^{N} z_i F C_i \quad (1)$$

$$\nabla \cdot \mathbf{J}_i = \nabla \cdot \left( C_i \mathbf{u} - D_i \nabla C_i - \frac{F z_i C_i D_i}{RT} \nabla \varphi \right) = 0 \tag{2}$$

$$\mu \nabla^2 \mathbf{u} - \nabla p - \sum_{i=1}^{N} (z_i F C_i) \nabla \varphi = 0 \tag{3}$$

$$\nabla \cdot \mathbf{u} = 0 \tag{4}$$

in which $\varphi$ and $N$ are the electrical potential and number of ion types. $F$, $R$, $T$, $\mu$, and $p$ are the Faraday's constant, gas constant, temperature, liquid viscosity, and pressure. $\varepsilon$ is the dielectric constant, and $\mathbf{u}$ is the fluid velocity. $\mathbf{J}_i$, $C_i$, $D_i$, and $z_i$ are the ionic flux, concentration, diffusion coefficient, and valence of ionic species $i$ (cations and anions), respectively. Note that in our finite element simulations, we cannot include the influence of ion size, ionic hydration, or ion-ion interaction on ion transport, which are usually considered in molecular dynamics simulations.[57-59]

In simulation cases, the radius and height of both reservoirs were chosen as 5 μm to avoid the reservoir size effect on the ion transport through nanopores.[60] The pore diameter ($D$) varied from 5 to 100 nm, and the pore length ($L$) changed from 10 to 200 nm.[14, 46, 61] $D$=10 nm and $L$=100 nm were selected as the default nanopore dimensions, which had been conveniently prepared in nanofluidic experiments.[62-64] The charged-length proportion($L_1/L$) of the inner pore surface was systematically varied from 0.01 to 0.99 to explore its influences on ionic current behaviors. The surface charge density and salt concentration varied from −0.005 to −0.12 Cm$^{-2}$, referring to the surface charge density of SiN, SiO$_2$, and polymer membranes,[48, 61, 65] and from 10 to 1000 mM, with the default values as −0.08 Cm$^{-2}$ and 100 mM, respectively. Four different electrolyte solutions, including KCl, LiCl, NaCl, and KF, were involved to consider the salt type effect. Diffusion coefficients of Li$^+$, Na$^+$, K$^+$, Cl$^-$ and F$^-$ ions were set to 1.03×10$^{-9}$, 1.33×10$^{-9}$, 1.96×10$^{-9}$, 2.03×10$^{-9}$, and 1.47×10$^{-9}$ m$^2$s$^{-1}$, respectively.[66] Solution temperature and relative permittivity were used as 298 K and 80. In this work, we focused on the ion transport through unipolar nanopores under electric fields. Simulation cases with concentration gradients[19-21, 67] across the nanopore will be conducted in the future study due to the additional heavy workload, where the coupled electric field and salt gradient should be considered.

As shown in Figure 1, the ground and working electrodes are placed in the reservoirs in contact with the neutral and charged nanopore ends, respectively. With the application of bias voltages ($V$) from −1 V to 1 V across the nanopores, corresponding ionic current values $I$ were obtained through ionic flux integration over the reservoir boundary with Eq. 5.[11, 30, 31]

$$I = \int_S F \left( \sum_i^2 z_i \mathbf{J}_i \right) \cdot \mathbf{n} \, dS \qquad (5)$$

where $S$ represents the reservoir boundary and $\mathbf{n}$ is the unit normal vector.

The applied mesh strategy, following our previous works, is shown in Figure S1.[14, 38, 46] A fine mesh size of 0.1 nm was selected on the inner and exterior (within 3 μm from the pore boundary) pore surfaces to consider the effect of EDLs on the ion transport through unipolar nanopores. A 0.5 nm mesh size was applied on the remaining exterior walls to reduce the calculation cost. Please note that for those simulation cases with charged exterior surfaces wider than 3 μm, mesh sizes of 0.1 nm and 0.5 nm were applied on the charged surfaces and the remaining surfaces, respectively. We have checked the reproducibility of our simulations by running an identical computational case on different computers. As shown in Figure S2 and Table S2, under the same boundary conditions, almost the same results were obtained with the same mesh strategy.

**Results and Discussion**

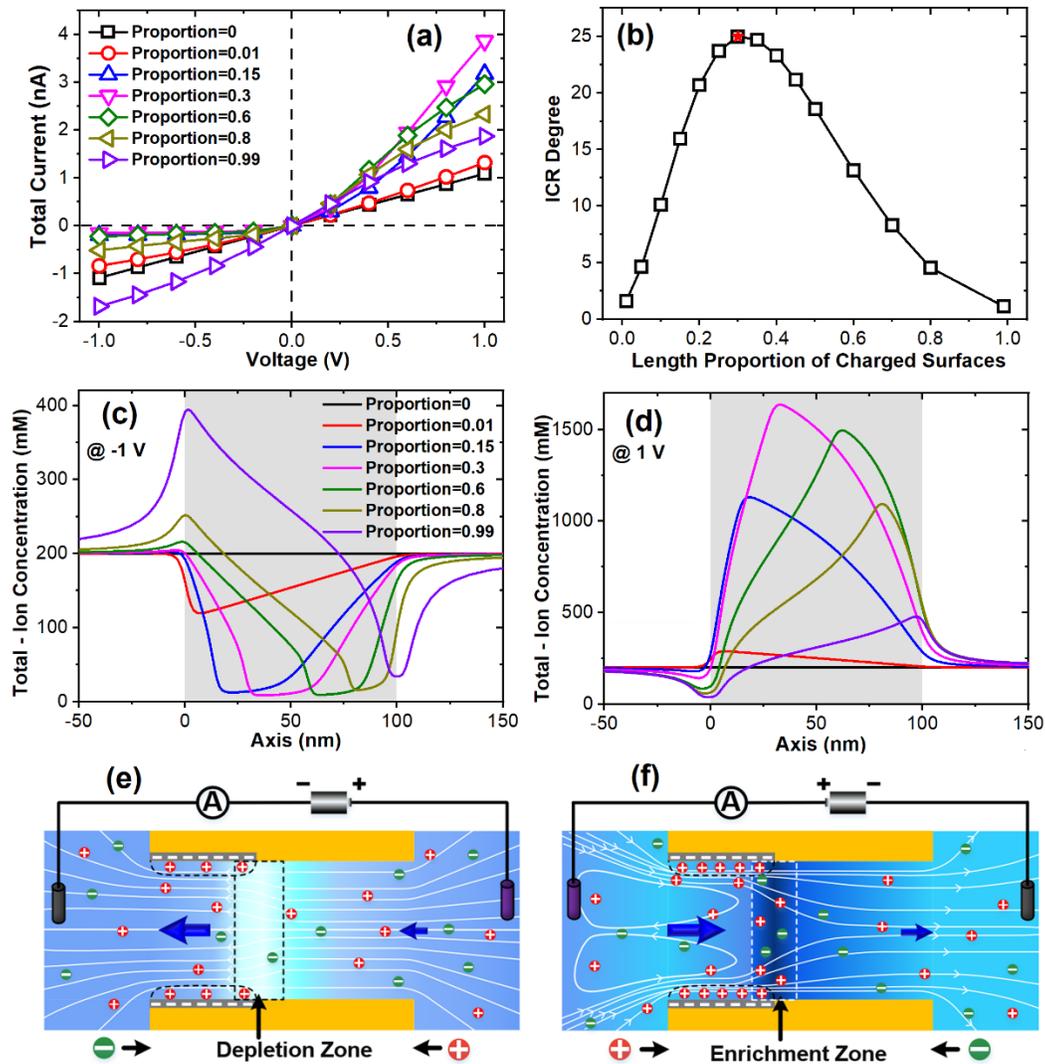

Figure 2 Ionic current behaviors through unipolar cylindrical nanopores with 100 nm in length under different length proportions of charged inner surfaces. (a) Current-voltage (I-V) curves. (b) Ionic current rectification (ICR) degrees in the nanopores at ±1 V. The red star shows the peak value. (c-d) Distributions of the total ion concentration along the pore axis at −1 V (c) and 1 V (d). The pore regions are shown in light grey. (e-f) Schemes for the formation of ionic depletion (e) and enrichment (f) zones at opposite biases. The pore length and diameter were 100 nm and 10 nm, respectively. Surface charge density was −0.08 C/m$^2$. Solutions were 0.1 M KCl.

In our simulations, short nanopores with 100 nm in length and 10 nm in diameter were considered, which can be easily realized on thin SiN membranes through micro/nanofabrication techniques.[6, 68] A series of unipolar nanopores has been considered by varying the charged-length proportion on inner pore walls (Figure 2). In nanofluidic experiments, the surface charge distribution in unipolar nanopores can be achieved through chemical modification.[28, 42, 43] Through exploring the effects of charged-length proportions on current-voltage (I-V) curves, the characteristics of ion transport in unipolar nanopores have been systematically investigated.

From Figure 2a, with neutral nanopores, linear I-V curves are obtained as predicted because of no obvious modulation of the ion concentration inside the nanopore (Figure 2c). Once a portion of the inner pore wall becomes charged, the asymmetric charge distribution on the inner surface significantly changes the ion transport, resulting in a nonlinear I-V curve known as ICR.[4] The larger and smaller ionic current under opposite voltage polarities corresponds to the "open" and "closed" states of diodes in electronics, reflecting a preferential direction of ion transport. At the charged-length proportion of 0.01, the ionic current is promoted and suppressed under positive and negative voltages, respectively, indicating that counterions predominantly enter or exit through the charged end. With the charged proportion increasing from 0.01 to 0.3, ionic conductance at the "open" and "closed" states is enhanced and inhibited gradually to their maximum and minimum. As the nanopore contains longer charged portions, ionic current at positive and negative voltages decreases and increases, respectively. When the charged length reaches 99% of the total inner surface, I-V curves exhibit an S-shaped profile with the appearance of limiting current at high voltages, which is due to the ion concentration polarization (ICP) across the nanopore caused by the strong ionic selectivity of the nanopore to counterions.[3]

The ICR degree is defined as the ratio of ionic current under positive and negative voltages through the unipolar nanopore. ICR degrees obtained at ±1 V are plotted in Figure 2b at different charged-length proportions. With the increase of the charged proportion, the ICR degree showed an increasing-decreasing profile. Different from the previous simulation work and understanding in long nanopores, the maximum ICR degree of ~25 appears at the charged-length proportion of 0.3 for 100-nm-long nanopores, instead of the typical 0.5 value observed in bipolar nanopores.[46]

ICR through unipolar nanopores has been studied in earlier works,[17, 26, 30] which originates directly from the ionic enrichment and depletion under opposite applied biases. Here, we further elucidate how the charged-length proportion modulates the ICR degree by analyzing ion concentration distributions along the nanopore axis, as shown in Figures 2c and 2d. From the schematic diagram (Figure 2e), under a voltage applied across the nanopore, free ions undergo electrophoretic movement inside the nanopore, where cations and anions move toward the cathode and anode, respectively. Due to the selectivity of the nanopore to counterions, the ionic movement inside the nanopore can be briefly divided into surface transport and bulk transport, i.e., ionic movement in the EDL region and the center region of the nanopore.[61, 69]

At negative voltages, cations enter and exit the nanopore through the surface transport along the charged segment and the bulk transport in the neutral region, respectively. Due to the enhanced cation and inhibited anion transport, the amount of free ions exiting the nanopore is much larger than that entering the pore, which induces a depletion zone inside the nanopore. The increased local resistivity in the depletion region causes a rise in local voltage. The induced stronger electric field accelerates the ionic electrophoretic transport out of the nanopore. Due to the positive feedback mechanism, the depletion zone reaches its equilibrium quickly. Under opposite voltages, more counterions are transported into the nanopore through the surface transport in EDL regions, resulting in ionic enrichment inside the nanopore at the location of the charge discontinuity. Because of the higher electrical conductivity in the ion-enriched region, a smaller partial voltage leads to a lower electric field strength. While in the vicinity of the charged part, cations are depleted at the entrance due to the fast ion transport. The induced larger electric field strength promotes the ionic migration further. Following a similar strategy to the formation of ion depletion, the positive feedback promotes the nanofluidic system to reach equilibrium. Based on our previous work,[38] the mechanism of ICR in unipolar nanopores is similar to that of the ICR in charged conical nanopores. However, the ICR degree of short unipolar nanopores is significantly higher than that of short conical nanopores.[14, 30, 33, 39] From our results in Figures 2c and 2d, significant ionic enrichment and depletion appear inside the pore, with the peak and valley values at the junction of charged and neutral regions.

Please note that near the charged regions, cations exhibit strong directional migration in EDL regions due to elevated local ion concentration. While in neutral pore regions, both cations and anions are transported through the central pore region. At positive and negative voltages, the formation of ion enrichment and depletion inside the nanopore can induce stronger ionic diffusion under concentration gradients, although this remains considerably weaker than the electrophoretic transport. Considering that the voltage is mainly applied across the nanopore, free ions enter the nanopore primarily by ionic diffusion under the formed concentration gradient between the inside and outside the nanopore.

For cases with longer charged regions, the ionic migration in EDL regions is greatly enhanced, which causes more significant ionic depletion and enrichment.[61] While the depletion/enrichment zone moves towards the neutral end of the nanopore, in this case, cations are more easily entering or exiting the pore. Both types of ion transport result in the most pronounced ion depletion/enrichment inside the pore at the charged proportion of ~0.3. Based on the concentration distribution along the pore axis, the pore has the highest and lowest electric resistance at ~0.3 simultaneously, instead of 0.5 as found in bipolar nanopores[46], which induces the most significant ICR.

Please note that when the proportion of the charged length on inner pore walls is greater than 0.6, ICP occurs across the nanopore due to the gradually enhanced ionic selectivity to counterions. With the rise of the applied voltage, the increase in the ionic current becomes slower gradually. When the charged proportion reaches 0.99, ICR disappears, and ICP becomes more serious, which leads to the limiting current in the I-V curve.[3, 14]

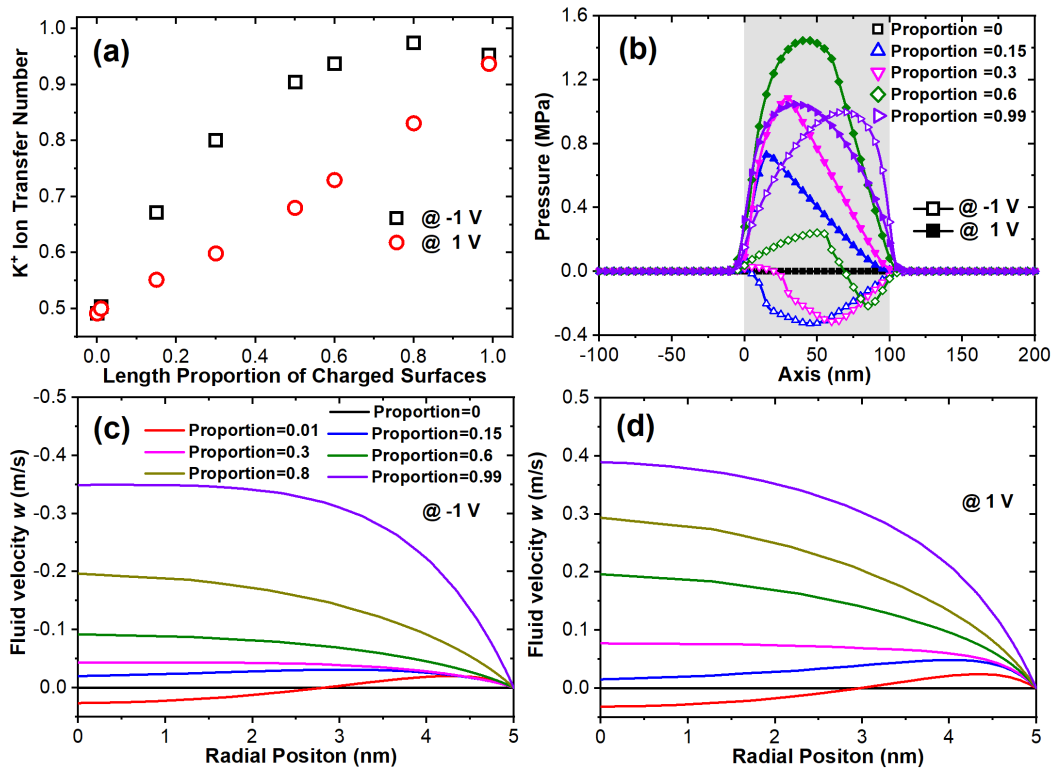

Figure 3 Ionic selectivity and fluidic behaviors inside unipolar nanopores under different length proportions of the charged inner-pore surface. (a) Transfer number of $K^+$ ions. (b) Distributions of induced pressure along the pore axis at ±1 V. (c-d) Radial distributions of electroosmotic flow (EOF) at ±1 V in the center cross-section of the charged regions.

Since the nanopore has an unipolar charge distribution, the nanopore is selective to the counterions.[70, 71] As shown in Figure 3, for the "open" state of unipolar nanopores, the ion selectivity depends linearly on the charged-length proportion. Because of the ionic hydration effect, the directional movement of counterions in EDLs results in electroosmotic flow (EOF).[72] The fluid flow provides a convective contribution to the ionic current. As shown in Figures 3c and 3d, due to the strong fluid flow near charged walls, EOF enhances the surface transport of ions in the EDL regions. This facilitates the entry and exit of ions through the nanopore.[73]

In neutral regions, without the enhancement of EOF, fluid flow is induced from the continuity of the flux. From Figure 3d, the stagnant liquid in the neutral region inhibits the fluid flow through the nanopore, resulting in a positive and negative hydrostatic pressure in the ion enrichment and depletion zones, respectively. The stagnant liquid facilitates the formation of ion enrichment and depletion.

Because a large number of counterions migrate rapidly along the EDLs into the nanopore from the charged inlet, the induced velocity of the fluid flow near the EDLs is higher than the flow velocity in the uncharged region. The flow hindrance beyond the EDL region induces the reverse fluid movement, i.e., along the axis towards the outside of the pore, which creates a vortex at the inlet. When the nanopore has charged proportions larger than ~0.6, the combination of the EOF and strong ionic selectivity causes more significant ICP, which dominates the modulation in ion concentration inside the nanopore (Figure 2d).[73]

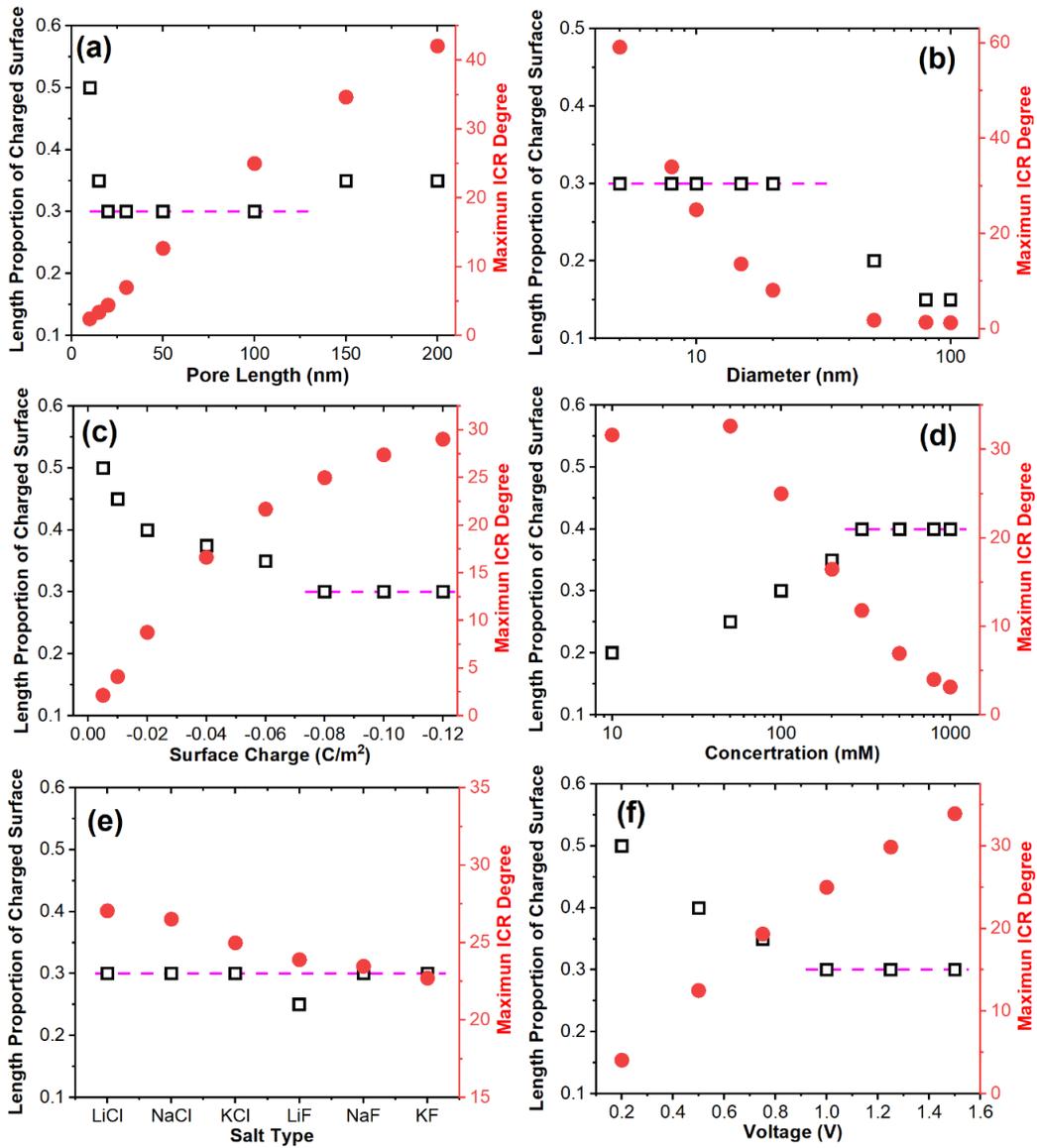

Figure 4 Maximum ionic current rectification (ICR) degrees (red dots) at ±1 V through unipolar nanopores, and the corresponding charged-length proportions (black squares) under various conditions, such as different pore lengths (a), diameters (b), surface charge densities (c), ionic concentrations (d), salt types (e), and applied voltages (f). The default pore length, diameter, surface charge density, solution, and voltage were 100 nm, 10 nm, −0.08 C/m$^2$, 0.1 M KCl, and ±1 V, respectively.

From the above statement, the largest ICR degree appears at the charged-length proportion of ~0.3. Here, we have investigated the dependence of the ICR degree and the corresponding charged-length proportion on the nanopore parameters and applied conditions.

I. Pore length. With the pore length varying from 10 to 100 nm, the obtained maximum ICR degree is enhanced linearly from ~1 to ~25 in 10-nm-diameter nanopores (Figure 4a), which is due to the more significant ion enrichment and depletion induced inside longer nanopores. While the charged-length proportion corresponding to the maximum ICR degree decreases from 0.5 to ~0.3, which approaches saturation when the pore length increases to 20 nm. In short nanopores, it is relatively easy for ions to diffuse into and out of the pore. With the pore length increasing, a longer charged wall is required to transport ions out of or into the pore in the EDL regions to form obvious ionic enrichment and depletion. As the pore length increases from 20 to 100 nm, the unipolar nanopores exhibit a maximum ICR degree at a proportion of ~0.3. This characteristic proportion becomes a property of unipolar nanopores that is independent of the considered pore sizes, surface charge density, voltages, and ion species. Please note that in our simulations, the step size for charged-length proportion variation is ~0.05. However, from Figure 4a and Figure S3, with the further increase of the pore length from 150 to 500 nm, the charged-length proportion reaches ~0.4.

II. Pore diameter. In a series of 100-nm-long nanopores, contrary to the effect of the pore length, with the nanopore diameter increasing, ICR in unipolar nanopores weakens, which presents a decreasing trend in the ICR degree (Figure 4b). As the diameter increases from 5 to 20 nm, the maximum ICR degree drops from ~60 to ~10. Inside the wider nanopores, the surface transport of ions plays a relatively weaker role in nanopore conductance.[14, 46] Due to the large aspect ratio, the charged-length proportion is maintained at ~0.3. When the pore size is larger than 50 nm, the EOF-induced ICP is enhanced, resulting in only ion depletion in the pores under opposite voltage polarities. The maximum ICR degree decreases further. At this time, the length ratio of the charged region where the maximum ICR degree occurs is ~0.2.

III. Surface charge density. Due to electrostatic attraction, surface charges attract counterions to form EDLs to shield the charged surface. As the surface charge density increases, more counterions are attracted to the wall. Under the action of the electric field, the electromigration of counterions becomes stronger, and the flow rate increases. It has a larger promotion on the bringing out and in of ions in the pores. Therefore, when the surface charge density is very low, the charged proportion is close to 0.5 (Figure 4c). As the charge density increases, the charged-length proportion decreases to ~0.3. With the increase in the charge density, the rectification phenomenon of the nanopore is more obvious, and the ICR degree increases monotonically.

IV. Salt concentration. The rectification mechanism of unipolar nanopores is similar to that of uniformly charged conical nanopores. In the previous study,[22] the rectification inside conical pores showed an increasing-decreasing trend with the concentration, reaching a maximum value at a concentration of ~10 mM. For the ultra-short 100 nm nanopores, a similar phenomenon is also found. As the solution concentration increased from 10 to 1000 mM, the ICR degree increased from ~30 to ~36 at 20 mM and decreased to ~2 (Figure 4d). The corresponding length proportion changes from 0.2 to 0.4. At low concentrations, the EDLs near the charged area are thicker, and the strong EOF promotes the transport of counterions in EDLs. Then, the maximum ICR degree is achieved at a short length proportion. As the ion concentration increases, thinner EDLs enhance the shielding effect of counterions on surface charges, which results in a longer charged surface reaching the maximum ICR degree due to the reduced electromigration of counterions.

V. Ionic species. Since negative charges were used in the simulations, the effect of the mobility of cations on the rectification properties inside unipolar nanopores was also considered. As shown in Figure 4e, the mobility of the counterions and coions has little effect on the maximum ICR degree. The charged-length proportion corresponding to the maximum ICR degree does not change. This is mainly attributed to that, during the movement of ions, the mobility is proportional to the diffusion coefficient,[74] the ions with a larger diffusion coefficient have stronger electromigration transport and diffusion motion simultaneously.

VI. Applied voltage. The applied voltage across the nanopore directly affects the electromigration of ions in the pore. As the voltage increases, the migration is enhanced, which in turn reduces the charged-length proportion required to reach the maximum ICR degree (Figure 4f). The ICR degree of unipolar nanopores increases with the voltage due to the more obvious ion enrichment and depletion inside nanopores caused by higher electric fields, as previously reported.[26, 27, 30]

Please note that here we focus on the exploration of optimal charged-length proportions for the best ICR performance, which may guide the design of unipolar nanopores.[47] In the literature, some experimental studies have shown that unipolar nanopores fabricated via chemical modification can present pronounced ICR. For example, Karnik et al.[16] introduced the surface charge discontinuity by selectively modifying half of a nanochannel with cationic avidin proteins while keeping the other half nearly neutral, and observed strong ICR through the channel. With single cigar-shaped polymer nanopores, Ali et al.[28] achieved uniform surface modification with pH-responsive carboxylic acid and lysine chains. Under a pH gradient of 2.5 and 8.5, their modified nanopore acted as a unipolar diode with an ICR degree larger than ~3. As shown by our simulations, the best ICR performance of unipolar nanopores depends on the charged-length proportions. However, in nanofluidic experiments, precise control over the surface charge properties of nanopores remains challenging.[42, 47] In the future, with the precise surface modification, high-performance unipolar diodes can be achieved conveniently.

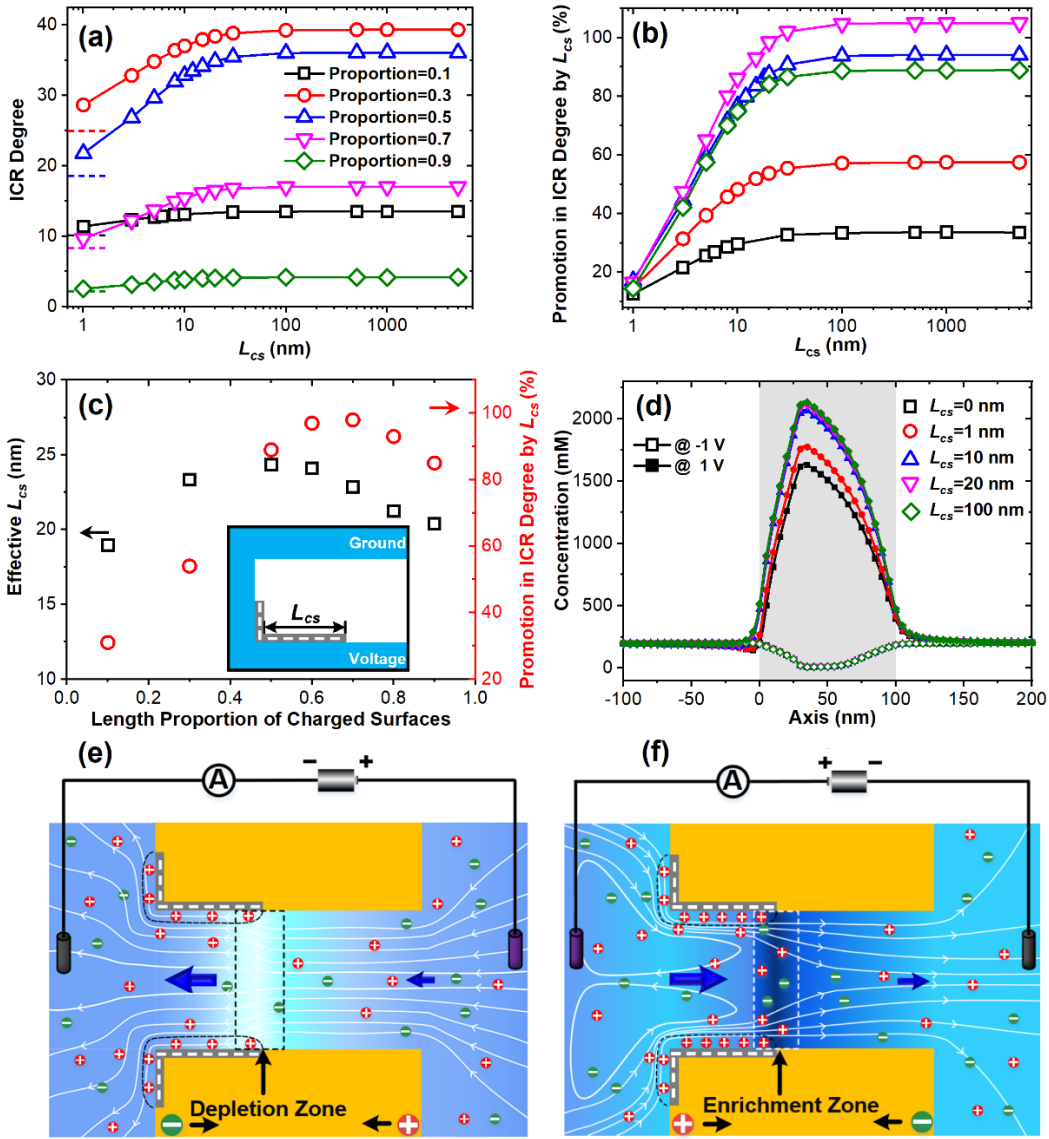

Figure 5 Ionic current rectification (ICR) degrees at ±1 V in unipolar nanopores with differently charged exterior surfaces ($L_{cs}$) on the charged inner-surface side. (a) ICR degree, (b) promotion in ICR degree by $L_{cs}$, and (c) effective $L_{cs}$ at different length proportions of charged surfaces. (d) Distributions of total ion concentrations along the pore axis under different $L_{cs}$ at ±1 V with a length proportion of charged surfaces at 0.3. (e-f) Schemes for the formation of ionic depletion (e) and enrichment (f) zones at opposite biases. The pore length and diameter were 100 nm and 10 nm, respectively. Surface charge density was −0.08 C/m$^2$. The solution was 0.1 M KCl.

For short nanopores, charged exterior surfaces have a significant impact on ion transport by providing fast passageways in EDL regions.[61, 69, 75, 76] Here, the influences of the exterior charged surfaces on ICR behaviors through unipolar nanopores have been systematically studied.[61, 69] The width of charged ring regions beyond the pore boundaries is defined as $L_{cs}$, as shown in the inset of Figure 5c. From Figure 5a, with the increase of $L_{cs}$, ICR degrees under different charged-length proportions increase first and then reach their saturation. Compared to the cases without exterior charges, fully charged exterior surfaces can enhance the ICR degree by at least ~33%. For the case with a charged proportion of 0.7, the enhancement can be ~105%.

As shown by Figures 5e and 5f, because of the exterior surface charges, the surface transport near the charged surface is enhanced significantly, which induces more obvious ion enrichment and depletion inside the nanopores. Normalized promotion in ICR degrees under various $L_{cs}$ is plotted in Figure 5b to the saturated value at $L_{cs}$ =5 μm. The effective $L_{cs}$ for each case is obtained when the normalized promotion reaches 95%, defined as $L_{cs\_eff}$. Please note that the 95% threshold was selected primarily due to the negligible error range in experiments of ~5%. Other values can also be considered as needed, such as 97% or 90%. Here, $L_{cs\_eff}$ ranges from ~20 to ~25 nm (Figure 5c), which shows similar values to our earlier work with conical nanopores.[61] The promotion in ICR by exterior surface charges presents an increasing-decreasing trend with the charged-length proportion. The largest promotion in ICR can reach ~100% when the charged-length proportion is ~0.7.

Distributions of the ion concentration along the pore axis are provided to show the modulation of ionic enrichment and depletion by exterior charged surfaces. From Figure 5d, ion enrichment increases with the larger charged area on the exterior surfaces. However, the charged area has little impact on the ionic depletion. From Figure S4, exterior surface charges have little influence on the ionic selectivity.

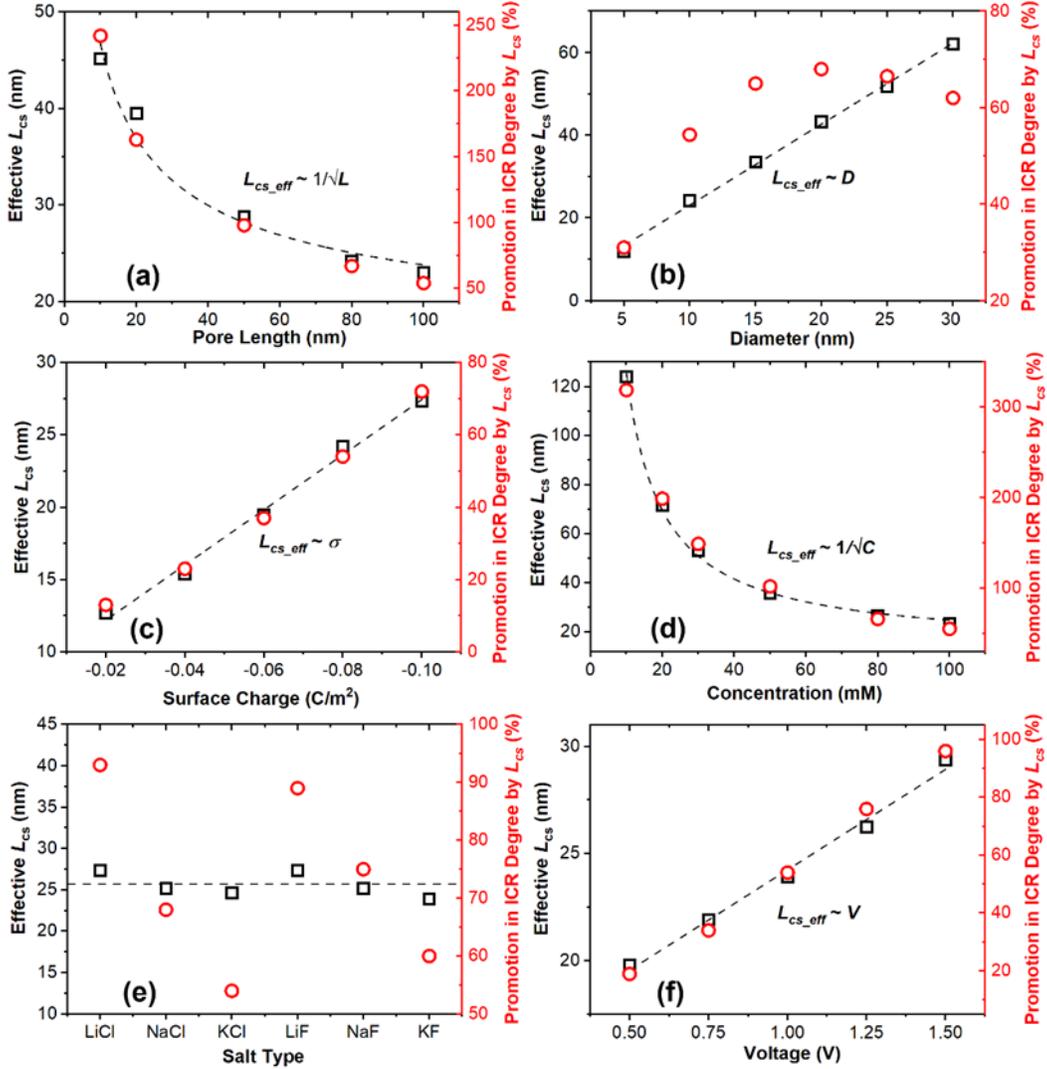

Figure 6 Effective width of the charged exterior surface ($L_{cs\_eff}$) and corresponding promotion in ionic current rectification (ICR) degrees for unipolar nanopores with a length proportion of charged surfaces at 0.3. Different nanopore properties and external conditions were considered, such as the pore length (a), diameter (b), surface charge density (c), ionic concentration (d), salt type (e), and applied voltage (f). Black dashed lines represent the fitting of $L_{cs\_eff}$ values. The default pore length, diameter, surface charge density, concentration, salt type, and voltage were 100 nm, 10 nm, −0.08 C/m$^2$, 0.1 M, KCl, and ±1V, respectively.

With a length proportion of charged surfaces at 0.3, the nanopore exhibits the strongest ICR response, which can be effectively amplified by the charge on the exterior walls. Here, we explored the effective width of the charged exterior surface ($L_{cs\_eff}$) beyond the pore orifice under different nanopore parameters and conditions with the charged-length proportion at 0.3. This work may provide help for the design of the porous membrane with unipolar diodes, where if the separation between adjacent nanopores is greater than $L_{cs\_eff}$, individual nanopores can function independently without any interference among nearby nanopores.

From the relevant nanofluidic experiments, charged exterior surfaces have been demonstrated to regulate the ion transport through nanopores.[75, 77, 78] Using silicon nitride pores with 700 nm in length, Nilsson et al.[79] achieved the surface charge modulation at the pore entrance by controlling oxide ring growth and chemical modification. As the length of modified exterior surfaces increased, the obtained current intensity through the nanopore enhanced. However, the conductance of experimental investigation can be extremely challenging for our simulated cases. So, we took the strategy developed by our group,[46, 61, 69] which has been verified by an experimental work of Yazda et al.[80] In their research, a minimum separation of ~500 nm was found for nanopores on a porous hBN/SiN membrane to work independently, which agreed well with the effective charged width of ~250 nm obtained in our previous osmotic energy conversion study with 30-nm-thick nanopores.[69]

From Figure 6a, the $L_{cs\_eff}$ is inversely proportional to the pore length. As the nanopore length increases from 10 to 100 nm, $L_{cs\_eff}$ gradually decreases from ~45 to ~23 nm, and the corresponding promotion in the ICR degree changes from ~242% to ~54% decreasing by ~77%. This is attributed to the increased pore resistance inside the longer nanopore, resulting in a smaller voltage drop at the pore entrance, which weakens the ion transport along the charged exterior surface. When the nanopore is sufficiently long, the impact of exterior surface charges on the ion transport is negligible.[14, 75]

In Figure 6b, $L_{cs\_eff}$ depends linearly on the pore diameter. As the pore diameter increases from 5 to 30 nm, the area of charged inner-pore surfaces expands, which accelerates the ion transport inside the EDL regions. Consequently, a larger $L_{cs\_eff}$ is required to provide sufficient counterions to the pore entrance along the charged external surfaces (Figure 6b). The corresponding promotion in the ICR degree presents an increasing-decreasing profile. For the nanopore with 100 nm in length, the pore diameter of 20 nm exhibits the largest promotion of ~68% in the ICR degree.

As shown in Figure 6c, $L_{cs\_eff}$ presents a linear relationship with the surface charge density. For nanopores with a higher surface charge density, more significant ion enrichment appears inside the nanopore under positive voltages, which induces a larger ionic current. The increased ion flux requires a larger charged exterior area to supply counterions from farther locations away from the orifices. The corresponding promotion in the ICR degree positively correlates with the surface charge density. As the surface charge density increases from −0.02 to −0.1 C/m$^2$, this promotion can be increased from ~13% to ~72%.

The solution concentration significantly affects the thickness of the EDLs inside the nanopore. At low concentrations, the thicker EDLs facilitate stronger ion transport, requiring a larger $L_{cs\_eff}$ to ensure sufficient ion supply. As salt concentration increases, the better screening of the surface charges from more counterions induces a weak modulation of ion transport by surface charges. Consistent with our previous findings in uniformly charged and bipolar nanopores, $L_{cs\_eff}$ in unipolar nanopores also exhibits an inverse relationship with salt concentration (Figure 6d).[46, 61] With the concentration increasing from 10 to 100 mM, the corresponding promotion in the ICR degree decreases from ~319% to ~55%.

The salt type has a negligible impact on the $L_{cs\_eff}$ despite the variations in diffusion coefficients. From Figure 6e, in solutions with different electrolytes, $L_{cs\_eff}$ remains almost at a constant value, ranging from ~23 to ~27 nm. This can be explained by the similar degrees of ion enrichment and depletion inside unipolar nanopores. However, due to the difference in the diffusion coefficients of cations in different electrolytes, $Li^+$ ions, with the smallest diffusion coefficient, present the most significant ICR degree enhancement of ~93%, followed by ~68% of $Na^+$ ions and ~54% of $K^+$ ions.

The dependence of the $L_{cs\_eff}$ on the applied voltage is exhibited in Figure 6f. In the system, applied voltages serve as the primary driving force for ion electromigration. Higher voltages can result in more significant ion enrichment, demanding a larger $L_{cs\_eff}$ to sustain the counterion transport along the charged external surface. As the voltage increases from 0.5 to 1.5 V, $L_{cs\_eff}$ increases from ~19.8 to ~29.4 nm. The corresponding promotion in the ICR degree increases from ~19% to ~96%.

**Conclusions**

In unipolar nanopores, the asymmetric distribution of surface charges on inner walls induces significant ICR. The charged-length proportion on inner walls can modulate the ion transport through nanopores, which produces various ICR performances due to the induced different degrees of ion enrichment and depletion. For a nanopore with 100 nm in length and 10 nm in diameter, the maximum ICR degree of ~25 is achieved at the charged-length proportion of ~0.3. Then, the charged-length proportion that yields the best ICR effect is found to be directly proportional to the pore length, surface charge density, and applied voltage, inversely proportional to pore diameter, and exhibits an increasing-decreasing trend with the salt concentration. These results provide design principles for unipolar nanopores with controlled ICR performance, which may find applications in nanofluidic sensors, ionic circuits, and ionic amplifiers.

For short nanopores, charged exterior surfaces can significantly promote ion transport by providing fast passageways in the EDL regions. Compared to cases without exterior charges, fully charged exterior surfaces enhance the ICR degree by at least ~33%. The effective width of the charged region on exterior surfaces increases linearly with the pore diameter, surface charge density, and applied voltage, while decreasing with the pore length and salt concentration, and remains independent of the salt type. These findings may guide the design of porous membranes with unipolar nanopores used in various applications such as parallel biosensing, blue energy harvesting, and seawater desalination.

## Supporting Information

See supplementary material for simulation details and additional simulation results.

## Acknowledgments

This research was supported by by the National Key R&D Program of China (2023YFF0717105), the Shandong Provincial Natural Science Foundation (ZR2024ME176), the Basic and Applied Basic Research Foundation of Guangdong Province (2025A1515010126), the Shenzhen Science and Technology Program (JCYJ20240813101159005), the National Natural Science Foundation of China (52105579), the Innovation Capability Enhancement Project of Technology-based Small and Medium-sized Enterprises of Shandong Province (2024TSGC0866), and the Key Research and Development Program of Yancheng City (BE2023010).